\documentstyle[multicol,aps,prl,floats,twocolumn]{revtex}

\draft

\begin{document}
\draft
\title{
{\tenrm\hfill Submitted to Phys. Rev. Lett.}\\
Inverse versus direct cascades in turbulent advection}
\author{M. Chertkov$^a$, I. Kolokolov$^b$ and M. Vergassola$^{c}$}
\address{$^{a}$ Physics Department, Princeton University, Princeton, NJ 08544, USA.\\
$^b$ Budker Institute of Nuclear Physics, Novosibirsk 630090, Russia.\\
$^c$ CNRS, Observatoire de Nice, B.P. 4229, 06304 Nice Cedex 4, France.}
\date{June 12, 1997}
\maketitle

\begin{abstract}
A model of scalar turbulent advection in compressible flow is analytically
investigated. It is shown that, depending on the dimensionality $d$ of space
and the degree of compressibility of the smooth advecting velocity field,
the cascade of the scalar is direct or inverse. If $d>4$, the cascade is
always direct. For small enough degree of compressibility, the cascade is
direct again. Otherwise it is inverse, i.e. very large scales are excited.
The dynamical hint for the direction of the cascade is the sign of the
Lyapunov exponent for particles separation. Positive Lyapunov exponents are
associated to direct cascade and Gaussianity at small scales. Negative
Lyapunov exponents lead to inverse cascade, Gaussianity at large scales and
strong intermittency at small scales.
\end{abstract}


\renewcommand{\thesection}{\arabic{section}}

The keystone of the celebrated 1941 Kolmogorov-Obukhov \cite{41Kol,41Obu}
theory for $3D$ fully developed turbulence is the {\em direct} (downscales)
energy cascade. Many other examples of direct cascades have later been found
for turbulent systems (see \cite{92ZLF,95Fri} for a review). The presence of
a direct cascade expresses the fact the average flux of an integral of
motion, which holds for the system \thinspace in the absence of forcing and
dissipation (e.g. energy for $3D$ Navier-Stokes turbulence and many examples
of wave turbulence, vorticity for $2D$ Navier-Stokes turbulence,
etc.),\thinspace is directed toward small scales and is constant along the
scales. The variety of turbulent systems is not exhausted by direct
cascades. When the pumping is supplied, the flux of an integral of motion
can go toward large, and not small, scales. This is the case of $2D$
turbulence, where the dynamical constraints due to the presence of two
integrals of motion lead to the remarkable phenomenon of the {\em inverse}
energy cascade discovered by R.H.~Kraichnan in \cite{67Kra}. Other examples
of inverse cascade are known in wave turbulence \cite{92ZLF}.

Advection of a passive scalar $\theta (t;{\bf r})$ (it might be the
concentration of a pollutant or temperature) by a turbulent {\em \
incompressible} flow belongs to the class of direct cascades \cite{48Obu}.
The direct cascade survives if the Navier-Stokes turbulent velocity is
replaced by a synthetic field with prescribed statistical properties \cite
{59Bat,68Kra} (this model, introduced by Kraichnan in \cite{68Kra}, is
attracting a great deal of attention for the anomalous scaling discovered
there \cite{95CFKLb,95GK,95SS}). A one dimensional compressible
generalization of the Kraichnan model was recently introduced in \cite{97VM}%
. We have considered the smooth limit of the model in \cite{97CKVa} and
shown that an inverse cascade takes place. This has led us to investigate
the general relation between compressibility and the direction of the
cascade. The aim of this letter is to present and analyze a model where we
can continuously move from inverse to direct cascade by varying two
parameters\,: the dimensionality of space $d$ and the degree of
compressibility.

The dynamical equation is 
\begin{equation}
\left( \partial _{t}\!+\!{\bf u}_{\alpha }(t;{\bf r})\nabla _{\alpha
}\!-\!\kappa \triangle \right) \theta (t;{\bf r})\!=\!\phi (t;{\bf {r}),}
\label{1}
\end{equation}
for the Lagrangian tracer scalar field $\theta (t;{\bf r})$ (say temperature
or entropy), steadily supplied by the random Gaussian source $\phi $ and
advected by the compressible $d$-dimensional Gaussian random velocity $u(t;%
{\bf r})$, having zero average. The correlation of the pumping is $\langle
\phi (t_{1};{\bf r}_{1})\phi (t_{2};{\bf r}_{2}\rangle =\chi ({\bf r}_{1}-%
{\bf r}_{2})\delta (t_{1}-t_{2})$, with $\chi ({\bf r})$ regular at the
origin and decaying fast at distances larger than the integral scale $L$.
The molecular diffusivity $\kappa $ is supposed small enough to be in the
fully turbulent regime, i.e. $L$ is much larger than the dissipative scale.
The pair correlation function of the velocity fluctuations, $\left\langle
\delta u_{\alpha }(t;{\bf r})\delta u_{\beta }(t^{\prime };{\bf r}%
)\right\rangle $, is 
\begin{equation}
\!\frac{2(dC^{2}\!-\!S^{2})r_{\alpha }r_{\beta
}\!-\!(2C^{2}-(d+1)S^{2})\delta _{\alpha \beta }r^{2}}{d(d-1)(d+2)}\delta
(t\!-\!t^{\prime }),  \label{correlation}
\end{equation}
where $C^{2}=\langle \left( \nabla _{\alpha }u_{\alpha }\right) ^{2}\rangle $%
, $S^{2}=\langle (\nabla _{\alpha }u_{\beta })^{2}\rangle $ and $\delta
u_{\alpha }(t;{\bf r})=u_{\alpha }(t;{\bf r})-u_{\alpha }(t;{\bf 0})$.

We show for the model that\,: {\em the cascade of the scalar is inverse if} $%
d<4$ {\em and the degree of compressibility} $C^{2}/S^{2} > d/4$\,; {\em %
otherwise it is direct}. The dimension $d=4$ turns then out to be critical
for the direction of the cascade. When the cascade is inverse and no
infrared cutoffs are present, the energy is clearly piled up at the large
scales and moments of the scalar field diverge linearly with time.
Correlations of scalar differences will however reach stationary values. The
PDF at the steady state of scalar differences is found explicitly in the
convective interval, i.e. at scales much larger than the diffusive one, both
for direct and inverse cascades. The Gaussianity of the scalar distribution,
known to be present at small scales in the incompressible case \cite{95CFKLa}%
, emerges at large scales when the cascade is inverse. Small scales are then
shown to be strongly intermittent.

Before the systematic analysis, it is worth presenting first some simple
intuitive arguments. Since the scalar is a tracer, its statistics is very
directly related to the one of the Lagrangian trajectories separation and,
specifically for smooth velocities, of Lyapunov exponents, describing the
rate of the exponential-in-time stretching (or contraction) of the
separation. We shall derive below the following relation between the sign of
the maximum Lyapunov exponent and the direction of the cascade: $\overline{%
\lambda }$ ${\em is}$ {\em positive in the case of direct cascade and
negative in the case of inverse cascade}. Its physical origin is simple to
explain. Indeed, once the statistics of the scalar field is expressed in
terms of Lagrangian paths properties, the crucial question becomes\thinspace
: given two particles initially separated by $r$, what is the probability
that their distance will ever reach the typical scale of the pumping $L$ ? 
{\em If the Lyapunov exponent is positive}, separations will typically grow
exponentially in time. Starting from a separation $r\ll L$, the scale of the
pumping is almost certainly reached ; viceversa, if $r\gg L$, there is
practically no chance to reach $L$. This is the dynamical hint of the direct
cascade. Consider now {\em a negative Lyapunov exponent}. The picture is
completely reversed. Since typical trajectories are contracting, the
probability to reach $L$ is much higher at scales $r\gg L$ than $r\ll L$.
The characteristic sign of the inverse cascade is emerging here.
Furthermore, the described physics of the relation between the
stretching-contraction interplay does have a certain generality. For
incompressible flow, the strain tensor being traceless, no strong trapping
phenomena are possible and the rate of typical separations growth is
expected to be positive. Strong trapping appear for compressible flow, where
it can lead to a substantial slow-down of transport (see \cite{VA}) and,
possibly, to a negative rate of the Lagrangian separations stretching.
Finally, the importance of traps for transport should definitely reduce when
the dimensionality of space increases.

Let us now start the systematic analysis of the model by considering
Lagrangian separations statistics. The stochastic differential equation
governing the evolution of the separation $R_{\alpha }(t)$ between two
Lagrangian particles, in the absence of molecular diffusion, is $\partial
_{t}R_{\alpha }=\sigma _{\alpha \beta }R_{\beta }$. The statistics of the
random strain $\sigma ^{\alpha \beta }$ is Gaussian. Its irreducible
correlation function $\langle \langle \sigma _{\gamma \alpha }(t)\sigma
_{\delta \beta }(t^{\prime })\rangle \rangle $ is simply obtained operating
with $\nabla _{{\bf r}}^{\gamma }\nabla _{{\bf r}}^{\delta }/2$ on (\ref
{correlation}). The average $\langle \sigma _{\alpha \beta }\rangle =-{\frac{%
1}{2d}}C^{2}\delta ^{\alpha \beta }$ is fixed by the condition $\langle
R_{\alpha }(t)\rangle =R_{\alpha }(0)$, which is an immediate consequence of
isotropy (see \cite{97CKVa} for more details). It is worth noting that the
average of $\sigma $ vanishes for incompressible flow. The $2n$-th moment of 
$R(t)$ obeys the following differential equation 
\begin{equation}
\partial _{t}\langle R^{2n}\rangle \!\!=\!\!{\frac{2n}{d}}\langle
R^{2n}\rangle \left[ S^{2}\left( \!{\frac{1}{2}}\!+\!{\frac{n-1}{d+2}}%
\right) \!+\!{\frac{2(n\!-\!1)}{d+2}}C^{2}\!\right] ,  \label{boh}
\end{equation}
corresponding to the following Gaussian statistics of the exponential
stretching rate $\lambda \equiv \ln [R(t)/R(0)]/t$\thinspace : 
\begin{eqnarray}
{\cal P}(t;\lambda ) &=&\sqrt{\frac{t}{2\pi \zeta ^{2}}}\exp \left[ -{\frac{%
\left( \lambda -\overline{\lambda }\right) ^{2}t}{2\zeta ^{2}}}\right]
,\qquad  \label{r2n} \\
\overline{\lambda } &=&\frac{dS^{2}-4C^{2}}{2d(d+2)},\qquad \zeta =\sqrt{%
\frac{S^{2}+2C^{2}}{d(d+2)}.}  \nonumber
\end{eqnarray}
The largest Lyapunov exponent (generally there are $d$ of them) has been
denoted by $\overline{\lambda }$ and $\zeta $ is the variance of $\lambda $.
For incompressible velocity fields, $C^{2}=0$ and the Lyapunov exponent $%
\overline{\lambda }$ is always positive. The opposite limit is the one of
gradient-type velocity fields ${\bf u}=\nabla \psi $, where the equality $%
C^{2}=S^{2}=\langle \left( \Delta \psi \right) ^{2}\rangle $ holds. The
interesting conclusion arising from (\ref{r2n}) is that $d=4$ is actually a
critical dimension for gradient-type fields. For generic smooth flow, the
largest Lyapunov exponent is always positive for $d>4$ and its precise
behavior (including the value of the possible critical dimension $%
d_{c}=4C^{2}/S^{2}$) depends on the specific value of the degree of
compressibility $C^{2}/S^{2}$. It follows from (\ref{r2n}) that, if the
Lyapunov exponent is negative, the low-order moments of the Lagrangian
separation decay in time, but their high-order (e.g. integer moments) grow
exponentially. As first highlighted in \cite{85ZMRS}, this means that the
dynamics of Lagrangian separations is dominated by rare events and this is
the origin of the strong intermittency of the scalar at small scales which
will be evidenced below.

We proceed now with the derivation of the scalar statistics. The relation
between Lagrangian trajectories and the passive scalar field in (\ref{1}) is
very direct. The solution of the equation can indeed be presented as $\theta
(t;r)=\int\limits_{-\infty }^{t}dt^{\prime }\phi (t^{\prime };%
\bbox{\rho}(t^\prime))$, where the Lagrangian trajectory $\bbox{\rho}(t)$
satisfies $\dot{\bbox{\rho}}(t^{\prime })={\bf u}(t^{\prime };%
\bbox{\rho}(t^\prime))$ and $\bbox{\rho}(t)={\bf {r}\ }$. For the sake of
simplicity of the presentation, we have not taken into account molecular
diffusion (see, for example, \cite{97CKVa} for the detailed path integral
formulation including diffusion). Its smearing effects on Lagrangian
trajectories can be neglected in the analysis presented in the sequel. The
simultaneous $2n$-th order scalar correlation function can be rewritten in
terms of the average over $2n$ Lagrangian trajectories as 
\begin{equation}
\langle \theta _{1}\cdots \theta _{2n}\rangle =\Biggl\langle \sum_{%
\mbox{permut.}}\prod\limits_{k=1}^{n}\int\limits_{0}^{T}dt\chi \left[ \frac{%
{\bf R}_{i_{k}j_{k}}(t)}{L}\right] \Biggr\rangle ,  \label{F}
\end{equation}
where the sum is over all the possible permutations $%
\{i_{1},...,i_{n},j_{1},...,j_{n}\}$ of the indices $\{1,\ldots ,2n\}$. The
Lagrangian separation ${\bf R}_{ij}(t)$ satisfies the Langevin equation
associated with (\ref{boh}) and the two Lagrangian particles are initially
located at ${\bf r}_{i}$ and ${\bf r}_{j}$, respectively. $T$ stands for the
time of evolution and should tend to infinity for the system to attain the
stationary state. The statistics of the whole set of $d$ Lyapunov exponents
is required to reconstruct the Lagrangian dynamics of a general $d$
dimensional structure. However, to analyze the structure functions of the
scalar (or multi-points correlation functions where all the points are
elongated on a straight line), we do not need to go into the heavy details
of subleading Lyapunov hierarchy. The point is that, modulo the dissipation
that will be discussed later on, the collinear geometry is preserved by the
dynamics \cite{95SS,95BCKL}. This issue actually deserves some more detailed
explanation. For the incompressible case, it has been found that the
transition from collinear to slightly off-collinear configurations is not
smooth \cite{95BCKL}. This implies, in particular, that the subleading
Lyapunov exponents statistics is needed to describe the correct rate of
exponential decay for the far tails of the scalar PDF \cite{97BGK}. Being
aware of this important effect, in the present letter we focus on the
quantities that can be analyzed by collinear considerations. The study of
the off-collinear effects is reserved for future publications. Here, we
shall therefore consider only collinear geometry ${\bf r}_{i}={\bf n}x_{i}$,
with all the distances large enough for dissipative effects to be
negligible. The $2n$-order structure function $S_{2n}(x)$ can be easily
reconstructed from the $2n$-th order correlation function $\langle \omega (T,%
{\bf r}_{1})\cdots \omega (T,{\bf r}_{2n})\rangle $ of the scalar gradient $%
\omega (t;{\bf r}_{i})\equiv \partial _{x_{i}}\theta (t;{\bf r}_{i}),$ in
the collinear geometry. This is simply done integrating all the coordinates $%
x_{i}$ from $0$ to $x$ and the final result for $S_{2n}=\langle (\theta
(x)-\theta (0))^{2n}\rangle $ is 
\begin{equation}
(2n-1)!!2^{n}\Biggl\langle \left[ \int\limits_{0}^{T}dt\left[ \chi
(e^{\lambda t}r_{d}/L)-\chi (e^{\lambda t}r/L)\right] \right] ^{n}\Biggr
\rangle,  \label{sf2}
\end{equation}
where the averaging with respect to $\lambda $ is fixed by (\ref{r2n}). The
regularity of the pumping $\chi (x)$ at the origin has been exploited in (%
\ref{sf2}) and the integrations over the $x_{i}$'s have been cut from below
by the diffusive scale $r_{d}=\sqrt{\kappa }$. We will indeed see that in
those cases when the cutoff is important, the resulting dependence on $r_{d}$
is logarithmic, thus justifying the used regularization. The generating
function ${\cal Z}(r/L,q)$ (Fourier transform of the respective PDF) of
scalar differences is simply restored from (\ref{sf2}) in the form of a
matrix element over an auxiliary quantum mechanics (see \cite{97CKVa} for
more details) with the Hamiltonian $\hat{H}=-\frac{\zeta ^{2}}{2}\partial
_{\eta }^{2}+q^{2}\left[ \chi (e^{\eta }/\mbox{Pe})-\chi (e^{\eta
}r/L)\right] $. Indeed, $Z=\left[ e^{-T\overline{\lambda }^{2}/(2\zeta
^{4})}\Psi (T;\eta )\right] _{\eta =0}$, where the wave function $\Psi $
satisfies the Schr\"{o}dinger equation $\left[ \partial _{T}+\hat{H}\right]
\Psi =0$, with the initial condition $\Psi (0;\eta )=e^{\overline{\lambda }%
\eta /\zeta ^{2}}$. The asymmetry between negative and positive $\overline{%
\lambda }$ clearly emerges in the different asymptotic behaviors of $\Psi $.
For negative $\overline{\lambda }$, the nonvanishing at $\eta \to -\infty $
initial condition $\exp [\overline{\lambda }\eta /\zeta ^{2}]$ will survive
in the stationary ($T\to \infty $) limit, since the potential part of the
Hamiltonian ( $\sim $ $q^{2}$) is also vanishing at $\eta \to -\infty $.
Analogous consideration of the positive ${\overline{\lambda }}$ case, fixes
the same boundary condition for $\Psi $, but in the opposite limit $\eta \to
+\infty $ (worth mentioning that a finite diffusivity is needed for this).
We thus arrive at the Fokker-Planck equation 
\begin{equation}
\left[ y^{2}\partial _{y}^{2}\!+\!\left( 1\!+\!\frac{2\overline{\lambda }}{%
\zeta ^{2}}\right) y\partial _{y}\!-\!\frac{2q^{2}}{\zeta ^{2}}\left[ \chi
(0)\!-\!\chi (y)\right] \right] {\cal Z}=\!0,  \label{FP3}
\end{equation}
with the following boundary conditions on ${\cal Z}(y,q)\!$ 
\begin{eqnarray}
&&\mbox{at}\quad \overline{\lambda }<0\quad {\cal Z}\Biggl|_{y\to 0}\to
1,\quad \left[ {\cal Z}y^{\overline{\lambda }/\zeta ^{2}}\right] \Biggl|%
_{y\to 1/\xi \to \infty }\to 0,  \label{min} \\
&&\mbox{at}\quad \overline{\lambda }>0\quad {\cal Z}\Biggl|_{y\to 1/\xi \to
\infty }\to 1,\quad \left[ {\cal Z}y^{\overline{\lambda }/\zeta ^{2}}\right] %
\Biggl|_{y\to 0}\to 0.  \label{max}
\end{eqnarray}
Here, $\xi $ should be considered as a $y$-independent parameter to be
replaced by $r_{d}/r$ after integration of (\ref{FP3}).

Note that the Fokker-Planck equation (\ref{FP3}) is the same as the one
found directly from (\ref{1}), taking two replicas, averaging with the
appropriate weight and discarding the contribution of the molecular
diffusivity term. If the latter term is kept, the so-called anomaly term
appears (following the field-theoretical terminology introduced by
A.~Polyakov in \cite{95Pol}) and the equation is not closed. In turbulent
systems, it is generally expected that the anomaly term, accounting for the
effects of molecular diffusion onto the convective interval range (for a
discussion of the Polyakov theory in the context of the present passive
scalar model, see \cite{97Yak}), does not vanish in the zero diffusivity
limit. The absence of the anomaly term is therefore neither a trivial nor a
general fact. Deriving (\ref{FP3}) dynamically within a multi-points
approach with an accurate fusion of points, we thus proved the absence of
anomaly in the present model.

Let us finally show how to get from (\ref{FP3}) and (\ref{min}) or (\ref{max}%
) the scalar differences PDF, with a special accent on the issue of
universality. The PDF is indeed not globally universal, in the sense that it
will generally depend on the explicit form of the pumping function $\chi (x)$%
. A relevant question to ask is then\,: what is universal in the
PDF\thinspace ? There are two different universality issues which may be
discussed in this respect. A kind of universality, typical of the case of
positive Lyapunov exponents both at large and small scales, suggests that
the only relevant parameter is the flux of $\theta ^{2}$ ($\chi (0)$) pumped
into the system at the integral scale \cite{95CFKLa}. Indeed, at $\overline{%
\lambda }>0$ the asymptotic solution of (\ref{FP3}) can be found replacing $%
\chi (0)-\chi (x/L)$ by $\chi (0)$. Using (\ref{max}) and the fact that $\xi 
$ is the smallest value in the problem, one gets an algebraic form for the
generating function ${\cal Z}=(r_{d}/\min \{r,L\})^{a}$, with the exponent $%
a=\sqrt{\overline{\lambda }^{2}/\zeta ^{4}+2q^{2}\chi [0]/\zeta ^{2}}-%
\overline{\lambda }/\zeta ^{2}$. Calculating the Fourier integral for the
PDF in the saddle-point manner (the large parameter is $r/r_{d}$), one gets
the Gaussian behavior at $|\delta \theta _{r}|\ll \ln [x^{*}]$%
\begin{equation}
{\cal P}(r/L,\delta \theta _{r})=\frac{\exp \left[ -\delta \theta _{r}^{2}%
\overline{\lambda }/(4\chi [0]\ln [x^{*}])\right] }{\sqrt{\overline{\lambda }%
/[\pi \chi [0]\ln [x^{*}]}},  \label{pos}
\end{equation}
where $x^{*}\equiv \min \{r,L\}/r_{d}$. For very large values $|\delta
\theta _{r}|\gg \ln [x^{*}]$, the Gaussian behavior transforms into an
exponential tail \cite{95CFKLa,94SS}. The detailed form of the tail requires
however the knowledge of the whole hierarchy of Lyapunov exponents and
cannot be calculated within the collinear approach used in the present
letter \cite{97BGK} at $d>2$. The unknown exponential tail does not affect
the behavior of the structure functions of order $n\ll \ln [x^{*}]$, that
are equal to $(2n-1)!!n!2^{n}\ln ^{n}[x^{*}]\chi ^{n}[0]/\overline{\lambda }%
^{n}$. This kind of universality (and the corresponding restrictions on
collinear considerations) can be easily extended to the case of negative
Lyapunov exponents for $r\gg L$. To this aim, we replace again $\chi
(0)-\chi (x/L)$ by $\chi (0)$ in (\ref{FP3}), impose the boundary condition (%
\ref{min}) and perform the saddle point integral, as it han been discussed
above. We arrive finally at the expressions for the PDF and the structure
functions (valid for $r\gg L$ and negative $\overline{\lambda }$), which are
identical to (\ref{pos}) with the replacements of $\min \{r,L\}/r_{d}$ by $%
r/L$ and $\overline{\lambda }$ by $|\overline{\lambda }|$.

For negative $\overline{\lambda }$, another kind of universal behavior is
found at scall scales $r\ll L$. In the vicinity of the origin, the PDF
depends only on the second derivative of the pumping $\chi ^{\prime \prime
}(0)$. This part of the PDF is indeed formed at the largest $q\gg 1$, where
one can keep only the first term of the expansion of $\chi (x/L)$ in the
``potential'' of (\ref{FP3}). We arrive then at the following expression for
the PDF at $|\delta \theta _{r}|\ll 1$ and $r\ll L$%
\begin{equation}
c\!\left( \frac{\chi ^{\prime \prime }[0]r^{2}}{\zeta ^{2}L^{2}}\right) ^{-%
\overline{\lambda }/\zeta ^{2}}\!\left[ \!\delta \theta _{r}^{2}\!+\!\frac{%
\chi ^{\prime \prime }[0]r^{2}}{\zeta ^{2}L^{2}}\!\right] ^{\overline{%
\lambda }/\zeta ^{2}-1/2},  \label{negdown}
\end{equation}
$c=\Gamma [1/2\!-\!\overline{\lambda }/\zeta ^{2}]/(\Gamma [-\!\overline{%
\lambda }/\zeta ^{2}]\sqrt{\pi })$. The expression (\ref{negdown}) shows
that all the structure functions $S_{n}(r)$ of low orders $-1<n<-2\overline{%
\lambda }/\zeta ^{2}$, are controlled by (\ref{negdown}) and scale normally $%
\sim r^{n}$. The behavior of the PDF at values larger than those where (\ref
{negdown}) holds, is not universal, i.e. it depends on the whole form of $%
\chi (x/L)$. For very large values $\delta \theta _{r}\gg 1$ and at small
scales $r\ll L$, it is however possible to show, from the expression of (\ref
{FP3}, \ref{min}) at $q\lesssim 1$ and $y\ll 1$, that the far tail is
exponential, ${\cal P}\sim \left[ r/L\right] ^{-2\overline{\lambda }/\zeta
^{2}}\exp (-\sqrt{\chi }\zeta |\delta \theta _{r}|/\overline{\lambda })$.
The non-universality emerges via the dimensional constant $\chi $, that
depends on the precise shape of the source function (for detailed
explanations on the asymptotics, see \cite{97CKVa}, where the tail was fully
determined in the one dimensional case for a specific form of the pumping).
The form of the tail, along with (\ref{negdown}), shows that the structure
functions of order larger than $-2\overline{\lambda }/\zeta ^{2}$ all scale
in the same way $S_{2n}(r)\sim (r/L)^{-2\overline{\lambda }/\zeta ^{2}}$ at $%
r\ll L$. This collapse of the high-order structure functions scaling
exponents should be contrasted with the Gaussian normal scaling observed for
the direct cascade. These different scalings are an illustration of the
profoundly different behaviors that have been found for the model below and
above its threshold of compressibility where the direction of the cascade is
reversed.

Illuminating discussions with E.~Balkovsky, G.~Falkovich, U.~Frisch,
K.~Gawedzki, V.~Lebedev, A.~Polyakov, B.~Shraiman, P.L.~Sulem, and V.~Yakhot
are gratefully acknowledged. We have benefited from the stimulating
atmosphere of the IHES, Bures-sur-Yvette, where we participated to the
workshop on turbulence. This work was partly supported by a R. H. Dicke
fellowship and ONR/DARPA URI Grant N00014-92-J-1796 (MC), Russian Fund of
Fundamental Researches under grand 97-02-18483 (IK), and the GdR
``M\'{e}canique des Fluides G\'{e}ophysiques et Astrophysiques'' (MV).

\end{document}